\documentclass[12pt]{article}
\title{
Color Superconducting State of Quarks \\}
\author{B.O. Kerbikov\\ State Research
Center\\Institute of Theoretical and Experimental Physics, \\
Moscow, Russia}
 \date{}
  \newcommand{\be}{\begin{equation}}
\newcommand{\ee}{\end{equation}}
\def\la{\mathrel{\mathpalette\fun
<}} 
\def\fun#1#2{\lower3.6pt\vbox{\baselineskip0pt\lineskip.9pt
\ialign{$\mathsurround=0pt#1\hfil ##\hfil$\crcr#2\crcr\sim\crcr}}}

\newcommand{\vek}{\mbox{\boldmath${\rm k}$}}

\newcommand{\veB}{\mbox{\boldmath${\rm B}$}}

\newcommand{\lan}{\langle}
\newcommand{\ran}{\rangle}

\begin{document}
\maketitle

\begin{abstract}
An introductory review of physics of color superconducting state
of matter is presented. Comparison with superconductivity in
electron systems reveals difficulties involved in formulating
color superconductivity theory at moderately ultra-nuclear
density.
\end{abstract}

{\bf Contents }

\begin{enumerate}
\item Introduction
\item The phases of QCD
\item
Symmetries of color superconducting phase
\begin{description}
\item{3.1}
Two flavors
\item{3.2} Three flavors
\end{description}
\item Dynamics of color superconducting phase
\begin{description}
\item{4.1} Comparing superconductivity in electron and quark
systems
\item{4.2} Gap equation
\item{4.3} Ginzburg-Landau free energy
\end{description}
\item Miscellaneous results
\begin{description}
\item{5.1} The ultra high density limit
\item{5.2} Rotated electromagnetism
\end{description}
\end{enumerate}

\section{Introduction}

During the last 2-3 years color superconductivity became a
compelling area of  QCD. The burst of interest to this topic was
triggered by papers \cite{1,2} though the subject has about two
decades history  \cite{3,4}. At present the number of publications
on color superconductivity is a two digit one. Our list  of
references includes only a minor part of them. The range of
questions related to the field became so wide that it can not be
elucidated and even touched in the present brief review. The
reader willing to get deeper and broader knowledge of the subject
may address review  papers \cite{5,6}.

The phenomenon of color superconductivity develops in the high
density regime of QCD when the  interaction of quarks starts to
feel the presence of the Fermi surface. Attractive quark
interaction makes the Fermi surface unstable to the formation of
the condensate of quark pairs (diquarks) with nontrivial color
structure. The term "color superconductivity" reflects the
similarity to  the behavior of electrons in ordinary
superconductors. The analogy is however not so complete as it will
be discussed in what follows.

In Section 2 we sketch the QCD phase diagram  in order to locate
the color superconductivity region in the density-temperature
coordinates.

In Section 3 the symmetries of color superconducting phase are
discussed. Section  4 is  devoted to the dynamics of the
condensate and to the  question to what extent the methods used in
Bardeen, Cooper and Schrieffer (BCS) theory are applicable.
Finally in Section 5 we present two unrelated topics chosen to
illustrate how rich the physics of color superconductivity is.

\section{The phases of QCD}
\setcounter{equation}{0} \def\theequation{2.\arabic{equation}}

It is well known that QCD in "normal" conditions (zero temperature
and density) still contains compelling open questions such as the
origin and dynamics of confinement. Despite important lacunas in
QCD at normal conditions substantial efforts have been undertaken
investigating QCD at nonzero temperature. These studies were
primarily motivated by heavy ion collision  experiments. Much less
attention has been devoted to QCD at nonzero density, i.e. to the
question "what happens to the matter as you squeeze it harder and
harder?" \cite{5}.

The reason why until recently high temperature region has obtained
much more attention than the high density one is twofold. As
already mentioned the development of QCD at $T>0$ was motivated by
heavy ion collision experiments while the nonzero density regime
is more difficult to reach in the laboratory since the temperature
in collisions is much hotter than the critical one for
superconductivity. In nature this state is realized in neutron
star interior which is beyond direct experimental probes. On the
theoretical side our main knowledge of $T>0$ QCD comes from
lattice Monte Carlo simulations. This powerful theoretical tool
does not work at nonzero density since in  this case the
determinant of the Dirac operator is complex resulting in
nonpositive measure of the corresponding path integral and the
failure of Monte Carlo simulation procedure. Recent breakthrough
in understanding of the high density QCD phase has been achieved
using models (like Nambu Jona-Lasinio or instanton gas) and
perturbation theory which is applicable in the limit of ultra high
density.

Before we start to draw the phase diagram the following remark is
in order. As we know from statistical mechanics  the actual
variable representing finite density in all equations in the
chemical potential $\mu$. For free massless quarks at $T=0$ the
connection between the density $n$ and the chemical potential
$\mu$ reads
\be
n=N_cN_f\frac{\mu^3}{3\pi^2}, \label{2.1} \ee where $N_c$ and
$N_f$ are  the number of quark colors and flavors respectively.

We begin the discussion of the phase diagram by reminding what
occurs along the vertical axis in the ($T,\mu)$ plane, i.e. at
$\mu=0$ \cite{5,7}. The central event happening along this line in
the transition from hadronic
 to the quark-gluon
phase. The hadronic  phase is characterized by the chiral symmetry
breaking and confinement. At the critical temperature which is
about 170 MeV \cite{8} a phase transition occurs, gluons and
quarks become deconfined and the chiral symmetry is restored. Such
a description of the temperature phase transition is a bird's-eye
view. One should keep in mind at least two remarks. First, the
fact that chiral symmetry restoration and deconfinement occur at
the same critical temperature is confirmed by Monte Carlo
simulations but not yet rigorously proved. Second, the value of
the critical temperature and the order of the phase transition
depend upon the number of flavors and the quark constituent
masses.

Now let us turn to our main subject, namely what  happens when one
moves to the right along the horizontal $\mu$  axis of the
$(T,\mu)$ plane. The vacuum state of QCD ($T=0,\mu=0)$ is
characterized by the existence of the chiral quark condensate
$\lan \bar \psi_L\psi_R\ran$. With $\mu$ increasing we enter the
nuclear (hadron) matter phase. Normal nuclear density $n_0\simeq
10^{-3}$ GeV$^3$ corresponds to $\mu_0\simeq 0.3$ GeV. The
behavior of the chiral condensate at such densities depends on
whether the formation of two quark droplets simulating hadrons is
taken into account explicitly. If not, the chiral condensate
remains practically at its vacuum value \cite{9}. If yes, the
phase is broken into droplets (hadrons) within which the chiral
symmetry is restored surrounded by empty space with condensate at
its vacuum value \cite{1}. Another condensate which is present
both in the vacuum and hadron phases is the gluon one. Its value
decreases with increasing $\mu$ \cite{10}.

Increasing $\mu$ further one reaches the point $\mu_1$ where
diquark condensate is formed and color superconducting  phase
arises. The difference  $(\mu_1-\mu_0)$ is of the order of the QCD
scale $\Lambda_{QCD} \simeq 0.2$ GeV \cite{7}. The transition at
$\mu_1$ is of the first order.  The main signature of this
transition is the breaking of the color gauge group SU(3)$_c$. At
this point we remind that the statement that local gauge
invariance is spontaneously broken is a convenient fiction
reflecting the fact that derivations are performed within a
certain fixed gauge \cite{11}. The pattern of breaking differs for
2 and 3 flavors and this will be discussed in Section 3. The
situation with chiral symmetry as also quite different for $N_f=2$
and $N_f=3$. In the first case it is restored in superconducting
phase while in the second it remains broken due to a nontrivial
mixing of color and flavor variables (color-flavor locking). The
discussion of this point is also postponed till Section3. To
stress the difference between superconducting phase for $N_f=2$
and $N_f=3$ the first one got the name $2SC$ (two flavors
superconducting) while the second one is called CFL (color-flavor
locked).

Despite essential differences of superconducting phases for
$N_f=2$ and $N_f=3$ the transition from one sector to another  is
performed by the variation of  a single parameter. This  is the
mass of a strange  quark: at $m_s=\infty$ we have a world with two
flavors, while at $ m_s=0$ -- the one with three. The physical
value of the strange quark mass $m_s\simeq 0.15 $ GeV is in
between these two extremes. If one can dial $m_s$ increasing it at
fixed $\mu$, one finds a first order transition from CFL phase
into a 2SC one. This happens because $\lan us\ran$ and $\lan
ds\ran$ condensates gradually become smaller than $\lan ud\ran$.
On the contrary  at asymptotically high densities $(\mu \to
\infty)$ the system for any $m_s\neq \infty$ is certainly in the
CLF phase. At $\mu=0$ the chiral symmetry restoration at $T_c$
occurs via second order transition for $m_s>m_s^c$ $(N_f$=2
regime) and via the first order for $m_s<m_s^c(N_f=3$ regime). The
value of $m_s^c$
 is estimated from lattice calculations as half of the physical
 mass of the strange quark.

 The final observation concerns the existence of the tricritical
 point $E$. Consider the two-flavor case with zero mass $u$ and $d$
 quarks. Then the phase transition at $\mu=0$, $T=T_c$ is, as we
 already know, of the second order, while the transition to the
 2CS phase at $\mu=\mu_1$, $T=0$ is of the first order. This means
 that the phase diagram features a tricritical poit $E$ to which a
 first order line approaches from the large $\mu$ side and the
 second order line emanates towards lower values of $\mu$ \cite{7}.

 \section{Symmetries of color superconducting state}

 The appearance of the diquark condensate at nonzero density leads
 to drastic changes in symmetries characterizing the system.

 The pairing of the  two quarks embodies a new feature somewhat
 unfamiliar for particle physics but inherent for BCS
 superconductivity. This is the existence of anomalous averages of
 two creation or annihilation operators which must be zero in any
 state with a fixed number of particles. The microscopic
 Hamiltonian proportional to $\psi\psi\bar\psi\bar\psi$ which
 according to the Wick's theorem is factorized  as  $\lan \bar
 \psi\psi\ran\bar\psi\psi$ conserves the number of particles and
 doesn't lead to superconductivity.  Superconducting state
 corresponds to an alternative factorization of the form
 $\lan\psi\psi\ran\bar\psi\bar\psi$ where $\lan\psi\psi\ran$ is
 the anomalous average which is zero in the  normal state. Such
 factorization corresponds to off-diagonal long range order, a
 concept introduced by Yang \cite{12}. There is a connection
 between the "nonconservation" of the number of particles and the
 breaking of the local gauge invariance mentioned before. Both are
 artifacts of a certain way of description  -- the mean field
 Hamiltonian in the first case fixed gauge in the other. The
 physically significant quantity is not the product of field
 operators but modules  of the gap. Its phase is significant only
 if one deals with actually open systems like in Josephson effect.
 At this point one may ask a question "How far the analogy between
 delocalozed Cooper pairs and diquarks extends?". In terms of
 condensed matter physics the question is whether diquarks
 resemble  Cooper  pairs \cite{13} or delute gas of more compact
 Schafroth  pairs \cite{14}.
 In current literature the first option is taken almost for
 granted. In our view the situation is not so obvious and we shall
 return to this question in Section 4.

 After these general remarks we turn to concrete symmetries of the
 quark system with nonzero condensate. Pairs of quarks cannot be
 color singlets, they may be either in color triplet or sextet
 state. The basic single gluon exchange, which is the QCD "Coulomb
 force", is attractive in $\bar 3$ color channel with color "wave
 function" proportional
 to $\lambda^\alpha\lambda^\beta\varepsilon_{\alpha\beta\gamma}$,
 where $\lambda$ are the Gell-Mann matricies. Instanton interaction also leads to attraction
 in $\overline{3}$ channel \cite{2}. Interaction in the
 sextet channel is believed to be weaker or even repulsive. Thus
 the condensate  in $\bar 3$ channel picks the color direction which means that gauge
 symmetry is broken. The situation reminds spontaneous
 magnetization below Curie point. The breaking pattern turns out
 to be quite different for $N_f=2$ and $N_f=3$. Therefore  the two
 cases are considered separately in the next two subsections.

 \subsection{Two flavors}
\setcounter{equation}{0} \def\theequation{3.\arabic{equation}}

At $\mu=\mu_1$ the first order phase transition from hadronic to 2
SC phase occurs. At $\mu<\mu_1$ the system features the condensate
$\lan \bar \psi_R\psi_L\ran$ which breaks chiral symmetry. At
$\mu>\mu_1$ the color superconducting condensate energes which has
the following structure
 \be
 \Delta\propto\lan \psi^{\alpha i}_{L}\psi^{\beta
 j}_{L}\varepsilon_{ij}\varepsilon_{\alpha\beta 3}
  \ran\propto-\lan \psi^{\alpha i}_{R}\psi^{\beta j}_{R}\varepsilon_{ij}
  \varepsilon_{\alpha\beta 3}
  \ran.
 \label{3.1}
 \ee

Here $\varepsilon$ are antisymmetric tensors, $\alpha$ and $\beta$
-- color indices, $i$ and $j$ -flavor ones, indices $L,R$ are
Lorentz  indices.  Pairing in this condensate is among quarks with
the same helicities, the pairs are ($ud -du$) flavor singlets, and
therefore  the condensate does not break chiral symmetry.
Transition is of the first order and therefore the  two
condensates may compete within a certain interval of $\mu$.
However as it was recently shown \cite{15} this is not the case
 and as soon as diquark condensate is formed the chiral one is
 extinguished in the sense that its influence on the thermodynamic
 properties of the system (the critical temperature and the gap)
 becoms negligible. Resorting to the analogy from solid state
 physics one may compare chiral condensate with nonmagnetic
 impurities in ordinary superconductor. According to Anderson
 theorem \cite{16} such impurities do not alter the properties of
 superconductor in the first order in their concentration.

 The color wave function of the  condensate (\ref{3.1}) is
 proportional to $\varepsilon_{\alpha\beta 3}$. This means that
 the first two colors (say red and green) are paired while the
 third one (blue) "remains in cold".
The condensate (\ref{3.1}) is invariant under the $SU(2)$ subgroup
of color rotations which do not affect the third (blue) quark.
Thus the color gauge symmetry is broken down to $SU(3)_c\to
SU(2)_c$. This symmetry pattern implies that five of the eight
gluons acquire mass via Anderson-Higgs mechanism. One of the three
massless gluons is mixed with a photon. This phenomenon called
"rotated" electromagnetism \cite{17} will be discussed in Section
5.

To summarize, in the $2SC$ phase color symmetry is reduced, chiral
symmetry is restored, photon is mixed with one of the eight
original gluons.

\subsection{Three Flavors}

Consider now QCD with three flavors of massless quarks. The
condensate is approximately of the form
\be
 \Delta\propto\lan \psi^{\alpha i}_{L}\psi^{\beta
 j}_{L}\varepsilon_{ijA}\varepsilon_{\alpha\beta A}
  \ran=-\lan \psi^{\alpha i}_{R}\psi^{\beta j}_{R}\varepsilon_{ij A}
  \varepsilon_{\alpha\beta A}
  \ran.
 \label{3.2}
 \ee
Summation over the index $A$ links color and flavor. This is the
famous color-flavor locking suggested in \cite{18}. The condensate
(\ref{3.2}) is invariant under neither color nor left-handed
flavor or right-handed flavor separately. It remains invariant
only under global $SU(3)$ rotation, so that the symmetry breaking
pattern is $SU(3)_c\times SU(2)_L\times SU(2)_R\to SU(3)_{c+L+R}$.
Color-flavor locking has a direct analogy in condensed matter
physics. This is the so-called $~^3He-B$ phase of superfluid
helium 3. In this phase the two atoms forming the pair are in
spin-triplet $S=1$ state with orbital momentum $l=1$. The
corresponding condensate has the form
\be
\Delta\propto(-p_x+ip_y)\chi(S_z=+1)+p_z\chi(S_z=0)
+(p_x+ip_y)\chi(S_z=-1),\label{3.3} \ee where $\chi$ is the spin
wave function, and $\Delta$ is invariant under combined orbital
and spin rotation corresponding to the total momentum $j=0$.

The fact that chiral symmetry in the CFL phase is broken makes it
difficult to distinguish it from hypernuclear matter, i.e. hadron
phase made of quarks with 3 flavors. This observation is called
quark-hadron continuity \cite{19}. There is also pairing in
hypernuclear matter, this time in dibaryon channels
$\Lambda\Lambda, \Sigma \Sigma$ and $N\Xi$ resulting in
superfluidity phenomenon.

Complete breaking of color gauge group in CFL phase implies that
all eight gluons become massive. Again as in the $2SC$ case photon
combines with one of the gluons.

In nature the two quarks are light and the third, the strange one,
is of middle weight. As already mentioned at $\mu\to \infty$ the
CFL phase is realized. With $\mu$ decreasing, or with $m_s$
increasing at fixed $\mu$, there is a critical point at which
strange quark decouples and two-flavor chiral symmetry is
restored. This unlocking phase transition (from CFL to $2SC$ ) is
of the first order \cite{20}.

\section{Dynamics of color supercoducting phase.}

 From BCS theory of superconductivitity we know that this
 phenomena arises under the following three conditions

 \begin{description}
 \item(i) Attraction between particles
 \item(ii) Existence of the Fermi surface
 \item(iii) Interaction of the particles must be concentrated
 within a thin layer of momentum space around the Fermi  surface,
 i.e. the Debye frequency has to be much smaller than the chemical
 potential, $\omega_D\ll\mu$. This requirement may be called
 Thin Shell Condition (TSC).
 \end{description}

 All the three points listed above are in principle met by quark
 system at finite density. However one shoud keep in mind certain
 reservations and substantial distinctions from electrons in
 metal. We address these issues in the next section.

 \subsection{Comparing Superconductivity in Electron and Quark
 Systems}

 QCD tells us that quarks in $\bar 3$
color channel attract each other while QED gives Coulomb repulsion
between electrons. Effective electron attraction is, as we know, a
result of a complicated mechanism involving lattice phonons. The
relevant feature of this interaction is that it satisfies TSC,
i.e. involves only electrons with energies $\omega$ close to Fermi
energy $\mu$, $|\omega-\mu|<\omega_{D}\ll\mu$, where $\omega_{D}$
is the Debye energy. Under this condition Cooper pairs completely
overlap, $n\xi^3\sim(10^8-10^{10})\gg 1, $ where $n$ is the
density of electrons, $\xi$ is the Pippard coherence length which
measures the spatial extension of the pair wavefunction.

 According to Section 2, in  QCD with two massless flavors
 transition to superconducting 2SC phase occurs at $\mu\simeq 0.4$
 GeV \cite{21} which recalling Eq. (\ref{2.1}) corresponds to
 $n\simeq 1.3\cdot 10^{-2}$ GeV$^3$.
  Pippard coherence length may be estimated as
 $\xi\simeq 1/\pi\Delta$ \cite{22}, where
$\Delta$ - is the value of the gap, $\Delta\simeq 0.1$ GeV
\cite{5,21}. This yields $\xi \simeq 0.6$ fm which agrees with the
estimate $\xi \simeq 0.8$ fm given in \cite{23}. Thus in  the
"newly born" 2SC phase  $n\xi^3 \la 1$. The corresponding Debye
frequency (parameter $\Lambda$ of Ref. (\cite{21})) is only
$\omega_{D}\simeq 2\mu$ instead of $\omega_D\gg \mu$ in  BCS. In
solid  state physics the limit $n\xi^3\ll 1, \omega_D\simeq \mu$
is known as Schafroth regime \cite{14} of Bose condensation in the
delute gas limit. With $\xi\simeq 0.6 $ fm the BCS regime settles
from $\mu\simeq$ (200-300) GeV.

In a more formal way the role of TSC will be displayed in  the
next subsection. We shall see that TSC provides weak coupling
solution of the gap equation. This is due to the fact that under
TSC particles interact in two dimensions instead of three and the
sum $\sum_k\vek^{-2}$ logarithmically diverges at small momenta.
Weak coupling solution is the cornerstone of superconductivity
theory.

Now we turn to another distinction of color superconductivity from
the BCS picture. The distinguishing feature of ordinary
superconductor is that it is a perfect diamagnet, i.e. magnetic
field $\veB=0$ inside it.  This is the famous Meissner effect. On
the other hand from QCD we know that color-electric and
color-magnetic field are "frozen" into the vacuum in the form of
the gluon condensate. With the quark density increasing the gluon
condensate is expected to "burn out" with qualitative description
of this process still lacking. Some insight may be gained from the
dilaton model \cite{24} which suggests that at $\mu\simeq 0.4$ GeV
when the 2SC phase arises the gluon condensate decreases by
(10-20)\% from its vacuum value \cite{25}.

From Section 3 we know that in color superconducting phase at
least part of the gluon degrees of freedom becomes massive. It
means effective screening of low-frequency modes and therefore it
is reasonable to assume that the formation of diquark condensate
should lead to the decrease of the gluon condensate. This may be
considered as nonabelian Meissner effect \cite{10}. Alternative
way of reasoning is the following. Suppose one performs
calculation of color superconducting state neglecting gluon
condensate and obtains the value of the chemical potential $\mu_1$
from which the condensation starts. With gluon condensate included
the system will remain in color superconducting state only if this
is thermodynamically favorable, i.e. provided the energy gain due
to transition into superconducting state exceeds the gain due to
the formation of the gluon condensate. As it was shown on general
grounds in \cite{10} and later confirmed by model calculations in
\cite{26} the superconducting state which has appeared at the
values of $\mu$ just above $\mu_1$ is destroyed by gluon
condensate. This means that the transition into superconducting
state is shifted to higher values of the chemical potential than
those predicted by calculations presented up to now. Solution of
the complete problem with  the inclusion of finite $T,\mu$ and
nonperturbative gluon fields remains a challenge for future work.

To summarize, we can say that comparison of color and electron
superconductivities leads to rather surprising conclusions.
Namely, the onset of the superconducting phase at low values of
$\mu(\mu\simeq 0.4$ GeV for 2SC) becomes questionable. The reason
is twofold. First, unlike the situation in BCS theory the
parameter $n\xi^3$ is not large. Diquarks resemble Schafroth pairs
rather than Cooper ones. The existence of the weak coupling
solution of the gap equation is not obvious since interacting
quarks are not necessarily confined to the two-dimensional layer
around the Fermi surface. Another reservation concerns the role
played by nonperturbative gluon fields. Due to gluon condensate
color superconductor is not a color diamagnet. At moderately low
chemical potential transition to the superconducting phase may be
blocked by gluon condensate.

\subsection{Gap Equation}
\setcounter{equation}{0} \def\theequation{4.\arabic{equation}}

In previous sections our presentation was very qualitative. From
now on we  shall be more formal but still avoid complicated
equations to be found in the original papers. Again our starting
point will be the BCS theory. Its formulation in the form close to
particle physics language may be found e.g. in Refs.
\cite{27}-\cite{29}. One starts with partition function written as
\be
Z=\exp \{V_4\Omega (\Delta, \mu,T)\},\label{4.1}\ee where $V_4$ is
the 4-volume of the system, $\Delta\sim \lan \psi\psi\ran$ is the
condensate, or gap, $\Omega$ is what is called effective potential
in the language of field theory or thermodynamic potential in
statistical physics. In BCS  theory at $T=0$ the
frequency-momentum representation for $\Omega$ reads \cite{29}
\be
\Omega_e=\frac{\Delta^2}{4g^2}+ \frac{i}{(2\pi)^4} \int d\omega
d^3 p\ln
\{1-\frac{\Delta^2}{\omega^2-\xi^2+i\varepsilon}\},\label{4.2} \ee
where $g^2$ is the interaction-constant, $\xi=p^2/2m-\mu$. The
stationary point of $\Omega_e$, i.e. the condition
$\partial\Omega_e/\partial\Delta=0$ gives the gap equation
\be
1=\frac{g^2\mu^2}{2\pi^2}\int^{\omega_D}_0
\frac{d\xi}{\sqrt{\xi^2+\Delta^2}}.\label{4.3}\ee
In passing from (\ref{4.2}) to (\ref{4.3}) one integrates the
second term in (\ref{4.2}) over $\omega$ and then resorts to TSC
by presenting the momentum integral as
\be
\int \frac{d^3p}{(2\pi)^3} =\int d\xi N_e(\xi) \simeq N_e(0) \int
d\xi,\label{4.4}\ee where $N_e(0)$ is the density of states at the
Fermi surface:
\be
N_e(0)=\frac{1}{2\pi^2} \left(p^2\frac{dp}{d\xi}\right)_F=
\frac{mp_F}{2\pi^2}.\label{4.5} \ee

The famous weak-coupling solution of Eq. (\ref{4.3}) reads
\be
\Delta=2\omega_D\exp \left\{-\frac{2\pi^2}{g^2mp_F}\right\}.
\label{4.6}\ee

Now we turn to color superconductivity and consider the 2SC case.
The corresponding effective potential has been derived in
\cite{21} (see also \cite{15} for details) and reads
\be
\Omega_q=\frac{\Delta^2}{4g^2} -\frac12 tr\ln \left\{
1+\frac{\Delta^2\varphi\varphi^+}{\hat p_+\hat p_-}\right\},
\label{4.7}\ee where $tr$ implies summation over color, flavor and
Lorentz indices and for $T\neq 0$ integration over  $dp_4$ is
replaced by summation over fermionic  Matsubara modes $p_4 =(2n+1)
\pi T$. The operator $\varphi$ in (\ref{4.7}) has the form
\be
\varphi=\varepsilon_{\alpha\beta 3} \varepsilon_{ij}
C\gamma_5=\varepsilon_{\alpha\beta 3} \varepsilon_{ij}
\gamma_2\gamma_4\gamma_5,\label{4.9}\ee and $\hat
p=p_\mu\gamma_\mu,~~p_\pm =(p_k, p_4\pm i\mu)$.

Expressions (\ref{4.2}) and (\ref{4.7}) differ in two points.
First, quarks have additional degrees of freedom -color and
flavor. Traces over color and flavor indices result in additional
factors
\be
tr_c\varepsilon\varepsilon =tr
(\delta_{\alpha\beta}-\delta_{3\alpha}\delta_{3\beta}) =2,~~
tr_f\varepsilon\varepsilon= tr \delta_{ij} =2. \label{4.10} \ee
Second, electrons in (\ref{4.2}) are nonrelativistic particles
with mass $m$ while (\ref{4.7}) is written for massless
relativistic quarks. The density of states for massless quarks
with color  and flavor (see (\ref{4.10})) is
\be
N_q(0) =\frac{2\mu^2}{\pi^2}, \label{4.11} \ee instead of
(\ref{4.5}) for electrons. The propagator of the relativistic
quark has two poles corresponding to positive and negative
frequencies. These two poles show up in the trace over Lorentz
indices of the term  $(\hat p_+\hat p_-)^{-1}$ in (\ref{4.7}). One
has
\be
tr_L\frac{1}{\hat p_+\hat
p_-}=\frac{1}{p^2_4+(p-\mu)^2}+\frac{1}{p^2_4+(p+\mu)^2}.\label{4.12}
\ee The second term (antiparticles contribution) is usually
neglected \cite{5} since its  denominator is everywhere greater
than $\mu$.

The gap equation is again obtained from stationary point of
$\Omega_q$ by making use of TSC and (\ref{4.10}), (\ref{4.11}) and
(ref{4.12}). It has  the form
\be
1=\frac{8g^2\mu^2}{\pi^2}
\int^{\omega_D}_0\frac{d\xi}{\sqrt{\xi^2+\Delta^2}}.\label{4.13}
\ee the weak coupling solution reads
\be
\Delta=2\omega_D\exp
\left\{-\frac{\pi^2}{8g^2\mu^2}\right\}.\label{4.14} \ee In the
models like NJL the Debye frequency $\omega_D$ and the coupling
constant $g^2$ are fitted simultaneously but  no unique "standard"
solution exists so far \cite{30}. As an educated guess one may
take $g^2=2$ GeV$^{-2}, ~~ \omega_D=$0.8 GeV and $\mu=0.5$ GeV.
This yields
\be
\Delta\simeq 0.15 {\rm GeV}\label{4.15}\ee in perfect agreement
with numerical calculations performed in various models, see
\cite{5} for references to the original works.

We wish to remind what are the main approximations and assumptions
on which the naive estimate (\ref{4.15}) is based. In writing the
expression (\ref{4.7}) for $\Omega_q$ the contribution from the
chiral condensate $\lan \bar \psi\psi\ran$ was omitted. This step
is justified by model-independent Anderson theorem \cite{15}. The
neglect of antiparticle contribution is also a   reliable
approximation. The applicability of TSC in getting weak-coupling
solution (\ref{4.14}) is questionable at low $\mu$ values as
discussed at length in the previous section. The role of
nonperturbative gluons in (\ref{4.7}) was completely ignored and
this might be a serious flaw of all approaches to color
superconductivity developed up to now. As it was shown in\cite{10}
the contribution of the gluon  condensate to $\Omega_q$ is at low
values of $\mu$ very close to the contribution from diquark
condensate and the crucial point is which force will win. With
these considerations in mind the estimates (\ref{4.14}) -
(\ref{4.15}) do not seem exceedingly oversimplified since
complicated calculations within different models are still plagued
by the tacit neglect of the factors listed above.

\subsection{Ginzburg-Landau Free Energy}

Expansion of the thermodynamic potential $\Omega_q$ containing the
second and fourth order terms is called Ginzburg-Landau
Functional. The first order term vanishes because of stationarity
condition while the sixth order on is important  near the
tricritical point. Ginzburg-Landau functional describes the
behavior of the system in the vicinity of the critical
temperature.

For superconducting quark matter this functional was first written
down in Ref.\cite{4}. Since then several authors addressed the
subject, see  e.g. \cite{31,32,15}. In the simplest case of
homogeneous condensate when the gradient terms are absent one has
\be
\Omega_q=\Delta^2 N_q(0)\left( \frac{T-T_c}{T_c}\right)
+\Delta^4\frac{N_q(0)}{T_c^2}\frac{7\zeta(3)}{16 \pi^2}
.\label{4.16}\ee

The critical temperature $T_c$ is about half the gap, i.e.
$T_c\sim 50 MeV$ \cite{5}. Comparing this value with the critical
chemical potential $\mu_1\sim 400$ MeV one concludes that quark
matter is a high temperature superconductor \cite{5}. On the other
hand $T_c$ is too low to be easily accessible in heavy ion
collisions.

The simplicity of the expression (\ref{4.16}) for Ginzburg-Landau
functional should not cover some underlying problems. From BCS
theory we know that in the Ginzburg-Landau phenomenological
approach it is assumed that magnetic field varies slowly over the
coherence length which is $\xi\leq$ 1fm for color superconductor
(see Section 4.1). On the other hand we know that nonperturbative
QCD fields have the characteristic correlation length $T_g\sim
$0.2 fm \cite{33}. This is the distance at which the QCD string is
formed and the width of the string between quarks. Again, as in
Sections   4.1-4.2 we encounter an open question concerning the
role of nonperturbative gluons.

\section{Miscellaneous  results}
\setcounter{equation}{0} \def\theequation{5.\arabic{equation}}

 The
number of problems discussed in relation to color
supercondutctivity  phenomena is already very wide. Two of them
will be touched upon in this concluding Section of our short
introductory review. These are ultra-high density limit \cite{34}
and modified  electomagnetism \cite{17}.

\subsection{The Ulltra-High Density Limit}

In our previous discussion we paid attention to somewhat shaky
grounds on which the low $\mu$ color superconductivity theory is
based. The region of  low and moderate chemical potential is of
great interest since there are hopes that this is the regime in
which color superconductivity might be realized in heavy ion
collisions and in the interior of the neutron starts. From the
theoretical standpoint the ultra-high energy limit is certainly
more refined.

With the chemical potential increasing the Fermi momentum becomes
large and asymptotic freedom implies that the interaction between
quarks is weak. In this case the gap can be calculated in
perturbation theory. The resulting solution drastically differs
from the classical BCS result (\ref{4.6}) and the analogous
expression (\ref{4.14}) assumed in low density regime of color
superconductivity. We remind that (\ref{4.14}) is written in terms
of the coupling constant $g^2$ with the dimension $m^{-2}$. This
constant may be considered as induced by one-gluon-exchange with
gluon line becoming very hard resulting in four-fermion
interaction. Although this gluon-line  roughening mechanism is far
from being  clear \cite{35,36} one is tempted to extrapolate
(\ref{4.14}) for $\mu\to  \infty $ in terms of dimensionless QCD
coupling constant $g(QCD)$ as $\Delta \sim \{-c/g^2(QCD)\}$. The
correct answer however is $(N_c=3,~N_f=2)$ \cite{34,37,38}
\be
\Delta\simeq\frac{\mu}{g^5(QCD)} b\exp
\left(-\frac{3\pi^2}{\sqrt{2}g}\right), \label{4.17}\ee where
$b=512 \pi^4 2^{-1/3} (2/3)^{5/2}$. The different  $g$ dependence
in the exponents (\ref{4.14}) and (\ref{4.17}) is due to
relativistic and retardation effects, While (\ref{4.14}) is
derived for contact interaction, (\ref{4.17}) fully accounts for
the structure of the gluon propagator. The result (\ref{4.17}) is
based on the Debye screening of electric gluons and Landau damping
of magnetic ones. What has been neglected in (\ref{4.17}) is the
effect of vertex corrections and hence (\ref{4.17}) is reliable
only for $\mu\gg10^8$ MeV. With the account of the $g(QCD)$
behavior dictated by asymptotic freedom the result (\ref{4.17})
implies logarithmic growth of $\Delta$ as $\mu\to \infty$. This in
turn means that the strange quark mass becomes irrelevant and the
CFL phase becomes more favorable than the 2SC one. The derivation
of (\ref{4.17}) from the first principles  of the QCD is probably
the most significant achievement of the color superconductivity
theory.

\subsection{Rotated Electromagnetizm}

In Section 3 we have already mentioned how  the photon couples to
color superconducting matter \cite{17}, namely that the original
photon combines with one of the gluons. Consider the CFL phase.
Evidently the condensate (\ref{3.2}) breaks  electromagnetic gauge
invariance since pairing occurs between differently charged
quarks. It is straightforward to find a combination of charge and
color generators for which all quark pairs are neutral. This "new
charge" is \cite{17}
\be
\tilde Q=Q+\eta \lambda_8\label{4.18}\ee with $\eta=1/\sqrt{3}$.
>From  (\ref{4.18}) it is easy to see that  all quarks have integer
$\tilde Q$ charge. The massless photon is
\be
\tilde A_\mu=\frac{g a_\mu+\eta e G^8_\mu}{\sqrt{\eta^2 e^2
+g^2}}=\cos \alpha_0A_\mu+\sin \alpha_0G^*_\mu.\label{4.19}\ee

The mixing angle $\alpha_0\simeq  \eta e/g$  is small. The
magnetic field corresponding to $\tilde A$ experience the ordinary
Meissner effect.

\section{Acknowledgements}

This introductory review resulted from a report made by the author
at ITEP Wednesday seminar. The author is grateful to the
participants for questions and remarks. Special thanks are to
N.O.Agasian,  V.I.Shevchenko and Yu.A.Simonov for numerous
illuminating discussions and N.P.Igumnova for the help in
preparing the manuscript. Support from grants RFFI-00-02-17836,
RFFI-00-15-96786 and INAS CALL 2000 project \# 110 is gratefully
acknowledged.

\end{document}